\documentclass[preprint,12pt,authoryear]{elsarticle}

\usepackage{amssymb}
\usepackage{amsmath}

\usepackage{hyperref}

\usepackage{listings}
\usepackage{color}
\definecolor{dkgreen}{rgb}{0,0.6,0}
\definecolor{gray}{rgb}{0.5,0.5,0.5}
\definecolor{mauve}{rgb}{0.58,0,0.82}

\DeclareMathOperator{\Tr}{Tr}

\journal{Computer Physics Communications}

\begin{document}

\begin{frontmatter}



\title{Slurm: fluid particle-in-cell code for plasma modeling}


 \author[kul,mao]{V.~Olshevsky\corref{cor1}}
 \ead{vyacheslav.olshevsky@kuleuven.be}

 \author[kul]{F.~Bacchini}
 \ead{fabio.bacchini@kuleuven.be}

 \author[kul]{S.~Poedts}
 \ead{stefaan.poedts@kuleuven.be}

 \author[kul]{G.~Lapenta}
 \ead{giovanni.lapenta@kuleuven.be}

 \cortext[cor1]{Corresponding author}

 \address[kul]{Centre for mathematical Plasma Astrophysics (CmPA), KU Leuven.
				Celestijnenlaan 200B, bus 2400, 
				B-3001 Leuven, Belgium}
 \address[mao]{Main Astronomical Observatory, 
 				27 Akademika Zabolotnoho St. 03143, Kyiv, Ukraine}


\begin{abstract}
With the approach of exascale computing era, particle-based models are becoming the focus of research due to their excellent scalability.
We present a new code, Slurm, which implements the classic particle-in-cell algorithm for modeling magnetized fluids and plasmas.
It features particle volume evolution which damps the numerical finite grid instability, and allows modeling of key physical instabilities such as Kelvin-Helmholtz and Rayleigh-Taylor.
The magnetic field in Slurm is handled via the electromagnetic vector potential carried by particles.
Numerical diffusion of the magnetic flux is extremely low, and the solenoidality of the magnetic field is preserved to machine precision.
A double-linked list is used to carry particles, thus implementation of open boundary conditions is simple and efficient.
The code is written in C++ with OpenMP multi-threading, and has no external dependencies except for Boost.
It is easy to install and use on multi-core desktop computers as well as on large shared-memory machines. 
Slurm is an ideal tool for its primary goal, modeling of space weather events in the heliosphere.
This article walks the reader through the physical model, the algorithm, and all important details of implementation.
Ideally, after finishing this paper, the reader should be able to either use Slurm for solving the desired problem, or create a new fluid PIC code.
\end{abstract}


\end{frontmatter}


\section{Motivation}
\label{sec:intro}
Particle-in-cell (PIC) method was first proposed by \citet{Harlow:1964} for modeling compressed fluids.
PIC combines particles which follow material motion and carry conserved quantities such as mass and momentum, with a grid on which the equations of motion are solved.
Computational particles in PIC could be considered as a moving refined grid, hence the method is useful for modeling highly distorted flows and interface flows.
Early implementations were cumbersome (especially in treating boundary conditions), memory-hungry, and unable to treat certain physical problems such as, e.g., stagnating flows, which made the method obsolete.
However, in the last two decades the method got its second wind thanks to the Cloud-in-Cell (CIC) algorithm \citep{Langdon:Birdsall:1970} for kinetic modeling of plasmas, and the Material Point Method (MPM) applied in solid mechanics \citep{Sulsky:1995}.
It is common nowadays to treat PIC solely as a method for kinetic plasma modeling, therefore narrowing its scope only to CIC \citep{Brackbill:2005}.
Both MPM and CIC delivered impressive new results and are further developing; we refer the reader to the reviews of the applications of CIC in space plasma modeling \citep{Lapenta:2012}; MPM in geophysics \citep{Sulsky:etal:2007,Fatemizadeh:Moormann:2015}, engineering \citep{NME:NME910}, and 3D graphics \citep{Stomakhin:etal:2013}.

Motivated by the above success we proposed a new PIC algorithm for magnetized fluids \citep{Bacchini:2017} which clarified and simplified many aspects of the original FLIP-MHD algorithm by \citet{Brackbill:1991}.
This article discusses the realization of our algorithm, \href{https://bitbucket.org/volshevsky/slurm}{Slurm}, implemented in C++ with OpenMP multi-threading.
We will walk you through all steps of the typical computational cycle of Slurm, its initialization, and main solver loop, explaining, where appropriate, data structures, architecture and physical models used in the code.

\section{Physical model}
\subsection{MHD equations}       
Slurm is designed (but not limited) to numerically advance in time a discretized set of conventional MHD equations in Lagrangian formulation which describe the dynamics of plasmas or magnetized fluids
\begin{eqnarray}
\label{eq:continuity} \frac{d\rho}{dt} = - \rho \frac{\partial u_l}{\partial x_l}, \quad \quad \quad \quad \quad \quad \quad \quad\quad \quad \quad \quad \quad \quad \quad\quad \quad \quad \quad \quad \quad \quad \\
\nonumber \rho\frac{du_j}{dt} = -\frac{\partial p}{\partial x_j} + \rho g_j + \frac{\partial}{\partial x_i}\left[ B_i B_j - \frac{1}{2}B^2\delta_{ij} \right] + \quad \quad \quad \quad \quad \quad \quad \quad\\ 
\label{eq:momentum}  \frac{\partial}{\partial x_i}\left[ \mu\left(\frac{\partial u_j}{\partial x_i} +\frac{\partial u_j}{\partial x_i}\right) + \left( \mu_v - \frac{2}{3}\mu\right)\frac{\partial u_l}{\partial x_l}\delta_{ij} \right], \\ 
\label{eq:energy}  \rho\frac{de}{dt} = -p\frac{\partial u_l}{\partial x_l} + \mu\left( \frac{\partial u_j}{\partial x_i} +\frac{\partial u_j}{\partial x_i} - \frac{2}{3}\frac{\partial u_l}{\partial x_l}\delta_{ij} \right)^2 + \mu_v\left( \frac{\partial u_l}{\partial x_l} \right)^2 + \eta J^2,
\end{eqnarray}
where summation is implied over the repeating indices $i, j, l$, which denote vector components; $d/dt=\partial/\partial t + u_j \partial / \partial x_j$ is the convective derivative.
The plasma is described by density $\rho$, velocity $\mathbf{u}$, pressure $p$, internal energy per unit mass $e$, and magnetic field $\mathbf{B}$; $\mathbf{g}$ represents external forces (gravity); $\mu=\nu\rho$ is the dynamic shear viscosity and $\mu_v=\nu_v \rho$ is the dynamic bulk viscosity\footnote{the notation and terminology follows the excellent book by \citet{Kundu:Kohen:2012}}; $\eta$ is the resistivity, and $\mathbf{J}=\nabla\times\mathbf{B}$ is electric current; $\delta_{ij}$ is the Kronecker delta.

In most cases reported in this paper, the adiabatic equation of state is assumed
\begin{equation}
e = \frac{1}{\rho}\frac{p}{\gamma - 1},
\label{eq:EOS}
\end{equation}
where $\gamma$ is the predefined gas constant.

\subsection{Magnetic field equation}
To close the above system of MHD equations, a condition must be imposed on the evolution of $\mathbf{B}$.
As a proxy for magnetic field, Slurm uses electromagnetic potentials $\mathbf{A}$ and $\phi$, where
\begin{equation}
\label{eq:BfromA}
\mathbf{B}=\mathbf{\nabla}\times\mathbf{A}.
\end{equation}
Using vector potential as a proxy for magnetic field, on a staggered grid (Yee lattice), the $\nabla\cdot\mathbf{B}=0$ condition is satisfied to machine precision \citep{Bacchini:2017}.
The staggered grid also makes our method second-order accurate in space.
In Slurm, the magnetic field is specified on the grid's cell centers, and electromagnetic vector potential is given on cell vertices (nodes).

The evolution of $\mathbf{A}$ is governed by the equation
\begin{equation}
\label{eq:A}
\frac{d\mathbf{A}}{dt} = \mathbf{u}\times\mathbf{B} + \left( \mathbf{u}\cdot\mathbf{\nabla} \right)\mathbf{A} - \nabla\phi - \eta \mathbf{J},
\end{equation}
where $\phi$ can be chosen freely to satisfy a certain gauge condition.
Although it has no effect on the solenoidality of the magnetic field, appropriate measures have to be taken to preserve the gauge condition throughout the simulation.
\citet{Tricco:Price:2012,Tricco:Price:2016} have introduced the constrained hyperbolic divergence cleaning of $\nabla\cdot\mathbf{B}$ in smoothed particle hydrodynamics, which was used later by \citet{Stasyszyn:Elstner:2015} to clean the divergence of $\mathbf{A}$.
According to this strategy, the scalar potential $\phi$ is evolved as
\begin{equation}
\label{eq:phi}
\frac{d\phi}{dt} = -c_h^2\nabla\cdot\mathbf{A}-\frac{\sigma c_h}{h}\phi, 
\end{equation}
where $c_h$ is the fast MHD wave speed, $h$ is the smoothing length (particle size), and $\sigma\approx1$ is a constant.
This way, the Coulomb gauge $\nabla\cdot\mathbf{A}=0$ or any other initial gauge could be preserved throughout the simulation.

Interestingly, in two dimensions 
\begin{eqnarray}
B_x=\frac{\partial A_z}{\partial y}, \\
B_y=-\frac{\partial A_z}{\partial x},
\end{eqnarray}
and by definition
\begin{equation}
\label{eq:Az2D}
\frac{d A_z}{dt}=0,
\end{equation}
hence $A_z$ defined on a moving grid (particles), does not change over time \citep{Bacchini:2017}.
Therefore in 2D (with no out-of-plane $B$ field) the gauge condition is preserved to roundoff, and no divergence cleaning is necessary.

\subsection{Artificial viscosity}
The PIC method has very low numerical dissipation, and in many problems related to shock handling a numerical bulk viscosity $\mu_\upsilon$ is useful to stabilize the solution.
Among many possible options we had tested, we obtained satisfactory results with the Kuropatenko's form of artificial viscosity \citep{Kuropatenko:1966,Chandrasekhar:1961}.
This viscosity is non-zero only in grid cells for which $\nabla\cdot\mathbf{u}>0$.
As given by \citet{Caramana:1998},
\begin{equation}
  \mu_\upsilon = \rho\left[ c_2\frac{\gamma+1}{4} \left| \Delta\mathbf{u} \right| + \sqrt{c_2^2\left( \frac{\gamma+1}{4} \right)^2 \left( \Delta\mathbf{u} \right)^2 + c_1^2 c_s^2} \right] \left| \Delta\mathbf{u} \right|,
\end{equation}
where $\left|\Delta\mathbf{u}\right| = \left| \Delta u_x + \Delta u_y + \Delta u_z \right|$ is the velocity jump across the grid cell, $c_s = \sqrt{\gamma p / \rho}$ is the adiabatic sound speed, $c_1$ and $c_2$ are constants close to $1$.

\section{Main classes and data structures}
\label{sec:architecture}
Traditional implementations of PIC (or, rather, CIC) algorithms use conventional domain decomposition for parallelizing the computation which, obviously, can lead to strong processor load imbalance. 
Different strategies were proposed to overcome it, such as particle splitting/merging \citep{Beck:Frederiksen:2016} or grid refinement and temporal sub-stepping \citep{Innocenti:etal:2015CPC}.
Particle methods, instead, provide excellent opportunity for task or event based parallel approach. 
The latter has been discussed by, e.g., \citet{Karimabadi:etal:2005JCP}, but no productive implementation has been reported yet, to our knowledge.

Slurm embraces a task based approach in which all particle operations are split between a user-defined number of OpenMP threads.
Distributing tasks, not computational sub-domains, is advantageous in problems with strong density imbalances.
Particles and computational grid exchange information several times during each computational cycle, but otherwise grid operations and particle operations are independent.
Therefore two main classes, \lstinline|Grid| and \lstinline|ParticleManager|, were chosen to manage grid elements (cells and nodes) and particles, respectively. 

\subsection{Basic players: cells, nodes, and particles.}
\begin{figure}
\begin{centering}
\includegraphics[width=0.5\textwidth]{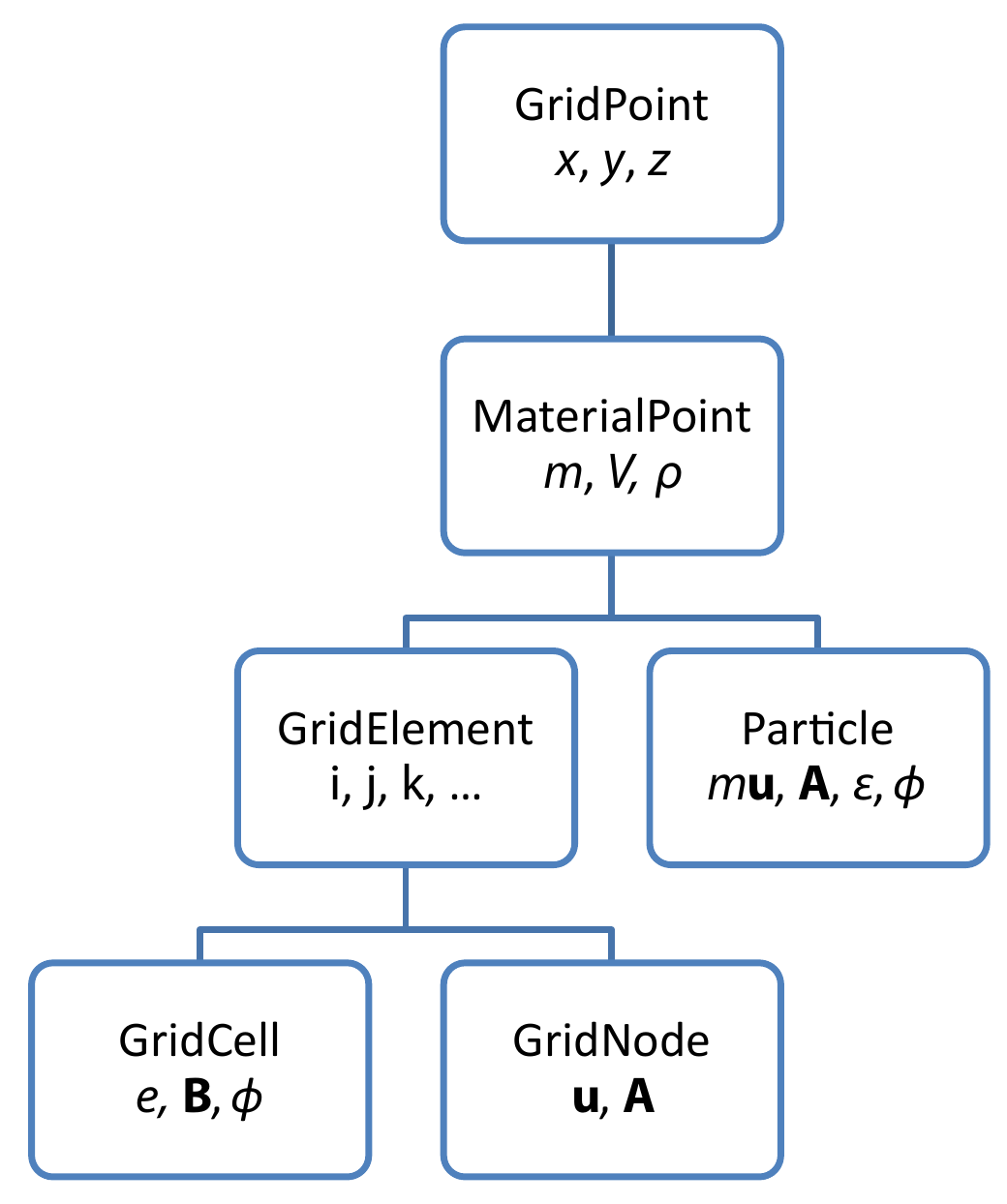}
\caption{Inheritance diagram for the basic entities in Slurm: points, grid elements and particles. Letters in each box indicate basic physical quantities carried by the corresponding item.
}
\label{fig:grid_elm}
\end{centering}
\end{figure}

We use a staggered grid, therefore three types of basic entities are represented by the \lstinline|GridCell|, \lstinline|GridNode|, and \lstinline|Particle| classes, which share some common properties as illustrated in Fig.~\ref{fig:grid_elm}.
Particles carry the following physical quantities: volume $V$, mass $m$, momentum $m\mathbf{u}$, internal energy $\epsilon=em$, and electromagnetic potentials $\mathbf{A}$ and $\phi$ (the latter is solely used to preserve the gauge of $\mathbf A$). 
Each grid element is assigned a mass $m$ and a volume $V$, hence the density is computed (upon interpolation from particles) as $\rho=m/V$, for both nodes and cells.
Grid nodes keep the fluid velocity $\mathbf{u}$ and the vector potential $\mathbf{A}$, while grid cells keep the internal energy density $e$, the magnetic field $\mathbf{B}$ and the scalar potential $\phi$.
In addition, all grid elements carry connectivity information as explained below.

\subsection{Grid}
\label{sub:grid}
The \lstinline|Grid| object is initialized by the solver first. 
Slurm supports rectilinear grids with $n_{xc}\times n_{yc}\times n_{zc}$ cells in each dimension, and $n_{xn}\times n_{yn}\times n_{zn}$ nodes in each dimension ($n_{x,y,z\,n} = n_{x,y,z\,c} + 1$). 
At each boundary, there is one layer of ghost cells, one layer of ghost nodes.
Each cell and each node stores information about its role (`general', `boundary', or `ghost') and pointers to its neighbors. 
Besides geometrical neighbors, logical neighbors are also computed.

Logical neighbors denote the grid elements which must be used for interpolating to/from this grid element.
For instance, a `logical neighbor node' of a boundary cell on a periodic boundary is the node on the opposite boundary.
The number of particles is much larger than the number of grid elements, hence connectivity information doesn't lead to excessive memory use.
However, it saves a substantial amount of resources and provides capabilities for implementing irregular grids.

\subsection{Particle Manager}.
\label{sub:PM}
Class \lstinline|ParticleManager| stores a double-linked list of \lstinline|Particle| objects, which makes adding and removing particles rather efficient and trivial to implement.
The list could easily be processed in parallel using OpenMP directives, e.g.,
\begin{lstlisting}
  #pragma omp parallel 
  {
    int thread_num = omp_get_thread_num();
    
    // Make sure particles were divided between threads according to the current number of threads
    if (omp_get_num_threads() != IteratorsNumberOfThreads) 
    {
      if (thread_num == 0)
        IteratorsNumberOfThreads = omp_get_num_threads();

      #pragma omp barrier
      computeOMPThreadRange();
    }

    // Process particles which belong to this thread.
    for(ParticleIterator pi = IteratorsBegin[thread_num]; pi != IteratorsEnd[thread_num]; ++pi)    
    {    
      // Do something
    }
  }  
\end{lstlisting}
where \lstinline|IteratorsNumberOfThreads| is the number of threads for which the particle ranges \lstinline|IteratorsBegin| and \lstinline|IteratorsEnd| were computed.
All expensive procedures: interpolation from particles to grid, interpolation from grid to particles, and particle push, are handled by the \lstinline|ParticleManager| object, and are parallelized in the above manner. 

\begin{figure}
\begin{centering}
\includegraphics[width=0.99\textwidth]{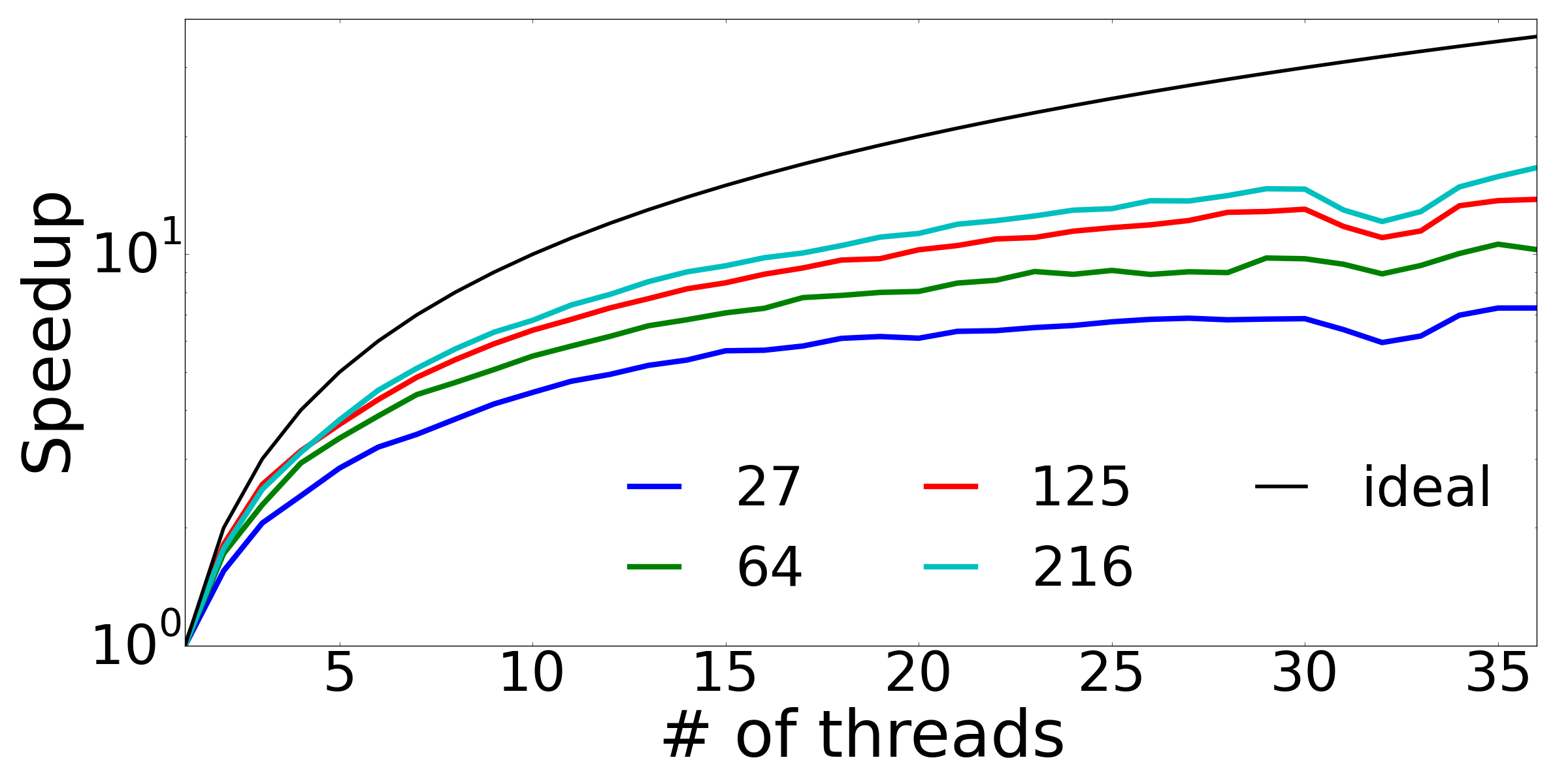}
\caption{Strong scaling on Intel Xeon E5-2600 v4 (Broadwell). The black line denotes ideal scaling; each color line corresponds to the number of particles per cell: from 27 to 216. The more particles per cell, the better the scaling is.
}
\label{fig:scaling}
\end{centering}
\end{figure}

Strong scaling tests performed on 36-core Intel Xeon processor E5-2600 v4 (Broadwell) have shown that parallel performance improves with the number of particles per cell (Figure~\ref{fig:scaling}).
To compute the parallel speedup, we ran simulations on the same $32^3$ grid, increasing the number of threads from $1$ to $36$.
The parallel speedup equals to the ratio of the average time of one computational cycle on one thread to the time on multiple threads $s(N)=\left<T(1)\right> / \left<T(N)\right>$.
On 36 threads the code executes 18 times faster than on a single thread, i.e., parallel efficiency reaches 50\%.
Note, the code has neither been optimized, nor tuned for performance. 
We used GNU C++ compiler 6.1 with the following options: \lstinline|-std=c++11 -O3 -fopenmp -lpthread|.

\section{Main solver loop}
\begin{figure}
\begin{centering}
\includegraphics[width=0.99\textwidth]{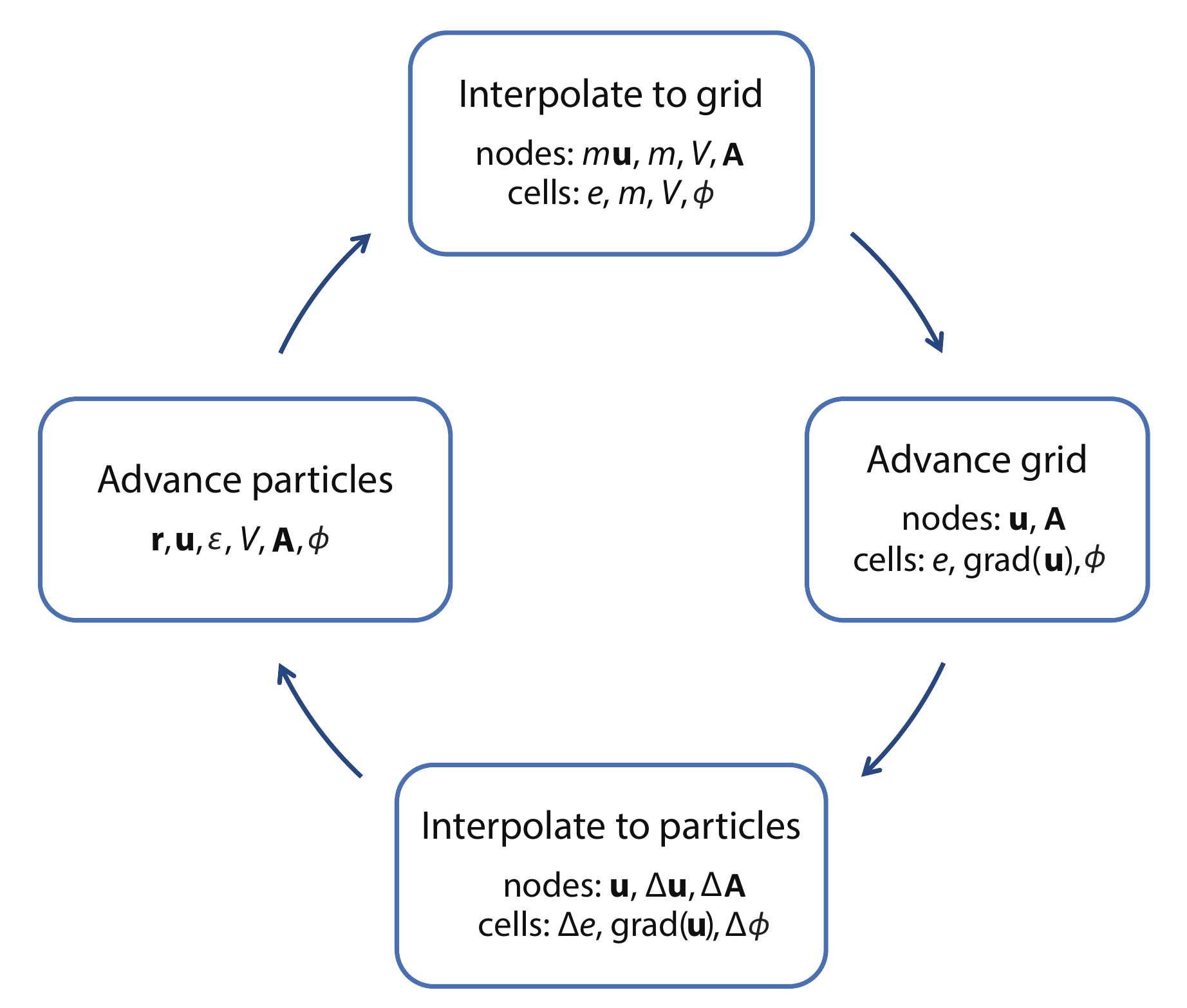}
\caption{Four-step explicit particle-in-cell algorithm implemented by Slurm.
}
\label{fig:algorithm}
\end{centering}
\end{figure}

The main solver loop of an explicit PIC algorithm consists of four steps: (1) interpolation particles$\rightarrow$grid, (2) grid advancement, (3) interpolation grid$\rightarrow$particles, and (4) particle mover, as illustrated in Figure~\ref{fig:algorithm}. 
In a typical PIC simulation the number of computational particles far exceeds the number of grid cells, and traversing all particles is the most time-consuming operation in the cycle.
In Slurm, particles are only traversed twice per cycle: first time in step (1) to interpolate to the grid. 
The second loop invokes both grid$\rightarrow$particle interpolation and particle push, therefore steps (3) and (4) are combined.
In the same loop particles are confronted against the boundary conditions, and are marked for deletion if needed.

\subsection{Step 1. Interpolation from particles to grid.}
\label{sec:interpolation}
Projection of data from particles to grid begins with setting the corresponding quantities ($m$, $V$, $e$, $\phi$ for cells and $m$, $V$, $\mathbf{u}$, $\mathbf{A}$ for nodes) to zero.
Interpolation is implemented in the \lstinline|ParticleManager| class in one loop over all particles.
It fetches the required information from each particle and adds the corresponding contribution to the surrounding grid nodes and grid cells.
When all particles have been processed, the accumulated values of the physical quantities on each grid element are normalized to the total interpolated weight at that grid element.

Information from each particle ($\epsilon$, $m$, $V$, $V\phi$) is projected onto 27 cells: the cell that encloses this particle, and 26 cells that share a face or a vertice with it.
Interpolation weights are given by the second order b-spline
\begin{equation}
w_{c,x} = \begin{cases}
            \frac{1}{2} x^2 - \frac{3}{2} \left| x \right| + \frac{9}{8} \quad \frac{1}{2} \leq \left| x \right| < \frac{3}{2} \\
            -x^2 + \frac{3}{4}       \quad\quad\quad\quad 0 \leq \left|x\right| < \frac{1}{2}
          \end{cases},
\end{equation}
where $x=\left(x_c - x_p\right) / \Delta x$ is the difference between the cell center coordinate $x_c$ and the particle's coordinate $x_p$ normalized to the cell's extent $\Delta x$.
The total weight is the product of the three weights in each dimension
\begin{equation}
w_c = w_{c,x} \cdot w_{c,y} \cdot w_{c,z}
\end{equation}

Conserved quantities to be projected on the nodes include $\mathbf{u}m$, $m$, $V$.
These are interpolated onto 8 vertices (nodes) of the cell which encloses this particle.
To compute interpolation weights, the conventional first order b-spline is used as described in \citet{Bacchini:2017}
\begin{equation}
w_n = \left(1 - \left|x_n - x_p\right|\right)\cdot\left(1 - \left|y_n - y_p\right|\right)\cdot\left(1 - \left|z_n - z_p\right|\right),
\end{equation}
where $x_n$, $y_n$, $z_n$ are the coordinates of this vertex (node).
A different procedure used to interpolate $\mathbf{A}$ from particles to grid.

Electromagnetic vector potential is not an additive conserved quantity, but rather a smooth vector function of three coordinates.
We interpolate $\mathbf{A}$ to each grid node from an arbitrary hexahedron formed by the node's closest particles.
Each grid node keeps information about the eight closest particles, hence the interpolation is realized in the \lstinline|Grid| class.
It is called when all conserved quantities have been projected.
Interpolation from irregular grids is quite straightforward in 2D\footnote{\href{https://www.particleincell.com/2012/quad-interpolation/}{https://www.particleincell.com/2012/quad-interpolation}}, but in three dimensions it is  more sophisticated, and requires the iterative solution of a nonlinear system.
For the sake of completeness, we provide the detailed description of the interpolation procedure adopted from NASA's b4wind User's Guide\footnote{\href{https://www.grc.nasa.gov/WWW/winddocs/utilities/b4wind_guide/trilinear.html}{https://www.grc.nasa.gov/WWW/winddocs/utilities/b4wind\_guide/trilinear.html}} in \ref{app:interpolation}.
Note, the same interpolation approach may be used for projecting the scalar potential $\phi$ on the cells, however none of our tests indicated that this is beneficial.

\subsection{Step 2. Advance the grid}
\subsubsection{Boundary conditions}
Boundary conditions should applied to all quantities interpolated from particles, and to all derived quantities such as $\mathbf{B}$ and $\mathbf{J}$.
All ghost and boundary grid cells and nodes carry the references to the applicable \lstinline|BoundaryCondition| objects.
They implement methods which modify the given cell or node quantities according to the specific rule.
For instance, in reflective boundary
\begin{lstlisting}
// class BoundaryConditionReflective : public BoundaryCondition
void BoundaryConditionReflective::imposeOnNode(GridNode * n, SlurmDouble t)
{
  for (SlurmInt i = 0; i < 3; i++)
    if (i == normal)
      n->set_u(normal, 0);
    else
      n->set_u(i, 2 * n->get_u(i));

  n->set_m(2 * n->get_m());
}
\end{lstlisting}

\subsubsection{Directional derivatives}
To evolve the equations of MHD in time, and to compute certain derived quantities such as $\mathbf{B}$, $\mathbf{J}$, etc., the gradients or directional derivatives of the discretized quantities should be computed.
In Slurm, a directional derivative of a node quantity $q$ in the direction $x$ is cell-centered.
It is computed as the average over the four corresponding edges of that cell.
Using three indices $\textrm{i}$, $\textrm{j}$, $\textrm{k}$, corresponding to $x$, $y$, $z$ coordinate dimensions, we designated the node-centered value as $q_{\textrm{ijk}}$, where each index is either $0$ for the `left' (bottom, front) or $1$ for the `right' (top, back) corner of the given cell.
Then the discrete directional derivatives are equal to
\begin{eqnarray}
\frac{\partial q}{\partial x} = \frac{q_{\textrm{100}} - q_{\textrm{000}} + q_{\textrm{110}} - q_{\textrm{010}} + q_{\textrm{101}} - q_{\textrm{001}} + q_{\textrm{111}} - q_{\textrm{011}}}{4 \Delta x}, \\
\frac{\partial q}{\partial y} = \frac{q_{\textrm{010}} - q_{\textrm{000}} + q_{\textrm{110}} - q_{\textrm{100}} + q_{\textrm{011}} - q_{\textrm{001}} + q_{\textrm{111}} - q_{\textrm{101}}}{4 \Delta y}, \\
\frac{\partial q}{\partial z} = \frac{q_{\textrm{001}} - q_{\textrm{000}} + q_{\textrm{101}} - q_{\textrm{100}} + q_{\textrm{011}} - q_{\textrm{010}} + q_{\textrm{111}} - q_{\textrm{110}}}{4 \Delta z}, 
\end{eqnarray}
where $\Delta x$, $\Delta y$, and $\Delta z$ are the cell's extents in the corresponding dimension.
The gradients of cell-centered quantities are node-centered.
They are computed in exactly the same way, except that $\Delta x$, $\Delta y$, and $\Delta z$ should represent the distances between the corresponding neighbor cell centers.
This way, the second derivative of a node quantity is node-centered, and the second derivative of a cell quantity is cell-centered.

\subsubsection{Solution of MHD equations}
A semi-implicit scheme is used to march the MHD equations on the grid in time.
First, from the cell-centered properties at the previous time step $\textrm{n}$, the new velocity is computed on each node.
\begin{equation}
\mathbf{u}^{\textrm{n+1}} = \mathbf{u}^{\textrm{n}} + \Delta t Q^{\textrm{n}},
\end{equation}
where $Q^{\textrm{n}}$ is the right-hand side of the discretized momentum equation \ref{eq:momentum} computed with the `old' grid values. 
Other grid properties are advanced using the new velocity according to discretized equations \ref{eq:continuity}, \ref{eq:energy}, \ref{eq:A}, \ref{eq:phi}, after the boundary conditions are imposed on $\mathbf{u}^{\textrm{n+1}}$
\begin{equation}
q^{\textrm{n+1}} = q^{\textrm{n}} + \Delta t Q\left(\mathbf{u}^{\textrm{n+1}}, \rho^{\textrm{n}}, p^{\textrm{n}}, \mathbf{A}^{\textrm{n}}, \phi^{\textrm{n}}\right),
\end{equation}
where $q$ is one of $\rho$, $e$, $\mathbf{A}$, or $\phi$, and $Q$ is the right-hand side of the corresponding MHD equation.
Note, boundary conditions need to be taken care of before interpolating the updated information to the particles.

The numerical scheme described above is a special case of the implicit scheme analyzed by \citet{Brackbill:Ruppel:1986}.
It is stable when the Courant condition is met, and offers better accuracy than a simple foward Euler method.

\subsection{Steps 3 and 4. Interpolate on, and push the particles}
The weights used to interpolate from grid to particles are the same as those used for particles to grid projection (Section~\ref{sec:interpolation}).
From grid cells to particles we interpolate the change of internal energy $\Delta \epsilon = \epsilon^{\textrm{n+1}} - \epsilon^{\textrm{n}}$, the cell-centered velocity gradient $\nabla\mathbf{u}^{\textrm{n+1}}$, and the change of scalar potential $\Delta \phi = \phi^{\textrm{n+1}} - \phi^{\textrm{n}}$. 
From grid nodes, the following quantities are interpolated: $\mathbf{u}^{\textrm{n+1}}$, $\Delta\mathbf{u} = \mathbf{u}^{\textrm{n+1}} - \mathbf{u}^{\textrm{n}}$, and $\Delta\mathbf{A} = \mathbf{A}^{\textrm{n+1}} - \mathbf{A}^{\textrm{n}}$.

Physical quantities carried by the $p-th$ particle are updated as
\begin{eqnarray}
\epsilon_p^{\textrm{n+1}} = \epsilon_p^{\textrm{n}} + m_p \Delta e, \\
\mathbf{u}_p^{\textrm{n+1}} = \mathbf{u}_p^{\textrm{n}} + \Delta\mathbf{u},\\
\mathbf{A}_p^{\textrm{n+1}} = \mathbf{A}_p^{\textrm{n}} + \Delta\mathbf{A}, \\
\phi_p^{\textrm{n+1}} = \phi_p^{\textrm{n}} + \Delta \phi.
\end{eqnarray}
The velocity gradient is used to advance the particle's volume according to either the strategy described by \citet{Bacchini:2017}, or in a simpler way:
\begin{equation}
V_p^{\textrm{n+1}} = V_p^{\textrm{n}} \left( 1 + \Delta t \Tr \nabla\mathbf{u}^{\textrm{n+1}} \right),
\end{equation}
where 
\begin{equation}
\Tr \nabla\mathbf{u}^{\textrm{n+1}} \equiv \nabla\cdot\mathbf{u}^{\textrm{n+1}} = \nabla\mathbf{u}^{\textrm{n+1}}_{11} + \nabla\mathbf{u}^{\textrm{n+1}}_{22} + \nabla\mathbf{u}^{\textrm{n+1}}_{33}
\end{equation}
is the trace of the interpolated velocity gradient tensor.
Note that particle quantities are updated using the interpolated changes in the grid quantities ($\Delta q$), which ensures very low numerical diffusivity of the method.
The interpolated velocity $\mathbf{u}^{\textrm{n+1}}$ is used only to advance the particle's position 
\begin{equation}
\mathbf{r}^{\textrm{n+1}} = \mathbf{r}^{\textrm{n}} + \mathbf{u}^{\textrm{n+1}} \Delta t.
\end{equation}
Once the new position is found, it is checked against boundary conditions.
If necessary, the particle is either reflected from the wall, flipped around the periodic boundary, or marked for deletion if it crosses an open boundary.

When all particles have been updated, the list of particles is walked through one more time, and all particles marked for deletion are removed.
Finally, if necessary, new particles are injected and the pre-computed particle ranges for each thread are updated.

\section{Examples}
Most testcases considered in this section could be directly compared with an excellent paper on Athena code validation by \citet{Stone:2008}.
Where appropriate, we provide references to the corresponding figures in this paper.

\subsection{Two interacting blast waves}
\begin{figure}
\begin{centering}
\includegraphics[width=0.99\textwidth]{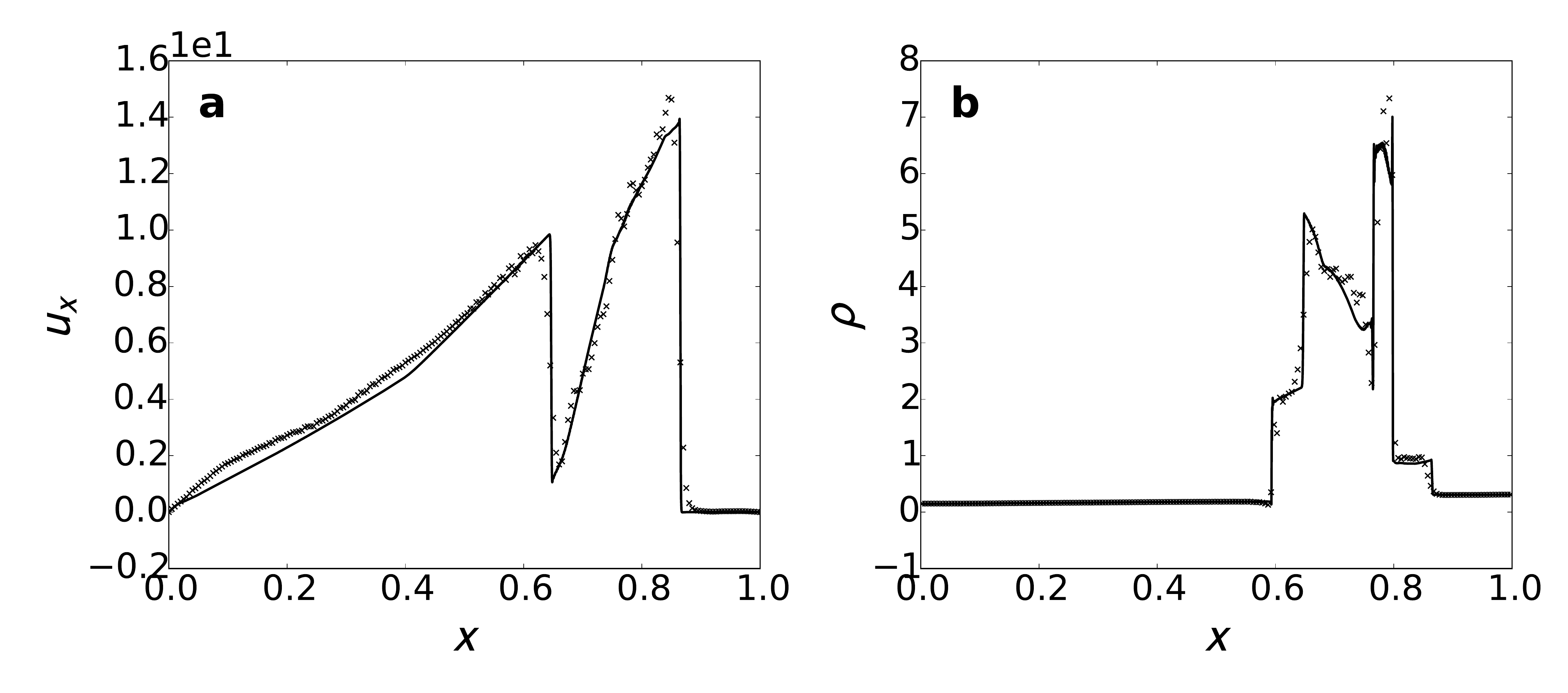}
\caption{Simulations of the two interacting blast waves at two resolutions, with $2400$ cells (solid line), and with $200$ cells (crosses), at $t=0.038$.
}
\label{fig:two-blasts}
\end{centering}
\end{figure}
A simple one-dimensional 1D shock tube proposed by \citet{Woodward:Colella:1984} is widely used to study the shock-capturing and stability properties of the codes.
In this problem, the initial state posesses a uniform density $\rho=1$, specific heats ratio $\gamma=1.4$, zero velocity, and three different pressures.
In the leftmost tenth of the domain $p=1000$, in the rightmost tenth $p=100$, and in between $p=0.01$.
The shock tube evolves quickly and produces multiple shock waves at high Mach numbers which are reflected from the walls and interact with each other.
This test is very hard to handle for Eulerian codes, however Slurm deals with it rather easy.
Step-by-step description of the complex evolution could be found in \citet{Woodward:Colella:1984}, we only present the final result at time $t=0.038$.

We performed two runs, with $2400$ and $200$ cells; both with $40$ particle/cell, and the same $\Delta t=4\cdot10^{-6}$.
The CFL condition computed on the fine grid using the largest initial value of the sound speed $c_s=\sqrt{\gamma p/\rho}=37.4$ suggests $\Delta t < 1.1\cdot 10^{-5}$.
The density and velocity profiles at $t=0.038$ in both simulations are shown in Figure~\ref{fig:two-blasts}.
These plots can be compared to the reference solutions in Figure~2h of the original paper \citep{Woodward:Colella:1984}.
Even at low grid resolution (200 cells), the contact discontinuity at $x=0.6$ (Fig.~\ref{fig:two-blasts}b) is very sharp.
However, even at high resolution there is an artificial spike at density discontinuity at $x=0.8$.
This spike neither grows, nor is a source of fictious oscillations, even when the simulation is run further.
More sophisticated forms of hyperviscosity might be needed to get rid of it in our simple explicit first-order numerical scheme.

\subsection{Brio \& Wu shock tube}
\begin{figure}
\begin{centering}
\includegraphics[width=0.99\textwidth]{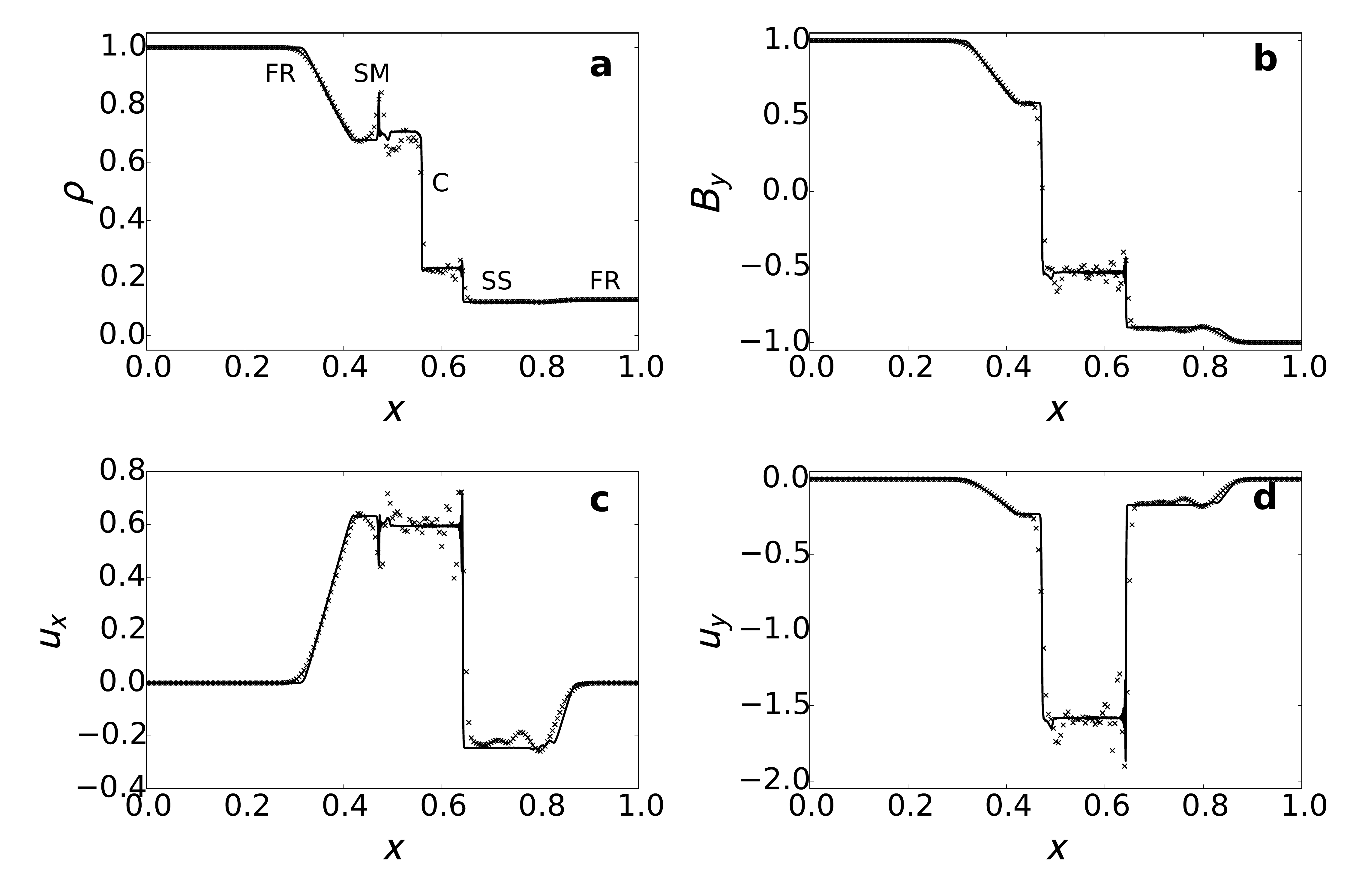}
\caption{Simulations of Brio \& Wu shock tube at two resolutions, with $3200$ cells (solid line), and with $400$ cells (crosses), at $t=0.1$.
}
\label{fig:brio-wu}
\end{centering}
\end{figure}
\citet{Brio:Wu:1988} proposed an extension of the Sod shock tube to MHD.
There are two initial states in the initial domain with $L_x=1$: the left state has $\rho=1$, $p=1$, and $B_y=1$; the right state has $\rho=0.125$, $p=0.1$, and $B_y=-1$.
In the whole domain, $B_x=0.75$, $\gamma=2$, and all other quantities are zero.
In terms of vector potential, the initial field is given by
\begin{equation}
A_z = \begin{cases}
            -x \quad\quad 0 \leq x < 0.5 \\
            x-1 \quad 0.5 \leq x \leq 1
          \end{cases}.
\end{equation}
At each computational cycle, when $\mathbf{B}$ is computed on grid cells from the interpolated $\mathbf{A}$ according to Eq.~\ref{eq:BfromA}, $B_x=0.75$ is added in each cell.

The aim of this test is to check how well the code distinguishes and handles different MHD shocks.
We present here the results of two runs: with $3200$ cells, $9$ particles/cell, $\Delta t=10^{-5}$, and with $400$ cells, $36$ particles/cell, $\Delta t=10^{-4}$.
In Figure~\ref{fig:brio-wu} the plots of $\rho$, $B_y$, $u_x$, and $u_y$ are shown at $t=0.1$.
These plots can be compared with Figure~2 of the original paper \citep{Brio:Wu:1988}.
Slurm accurately reproduces two fast rarefaction waves (FR), a slow compound wave (SM), a contact discontinuity (C), and the slow shock (SS).
There is still some numerical noise around the slow shock boundary, similar to a density spike in the interacting blast waves simulation.
This noise is efficiently damped by the Kuropatenko's hyper-viscosity, and is not affecting other cells of the grid.
It neither grows in amplitude, nor propagates further when the simulation runs well beyond the $t=0.1$ time.
In this simulation, high spatial resolution is essential to correctly handle the shocks.

\subsection{Orszag-Tang}
\begin{figure}
\begin{centering}
\includegraphics[width=0.99\textwidth]{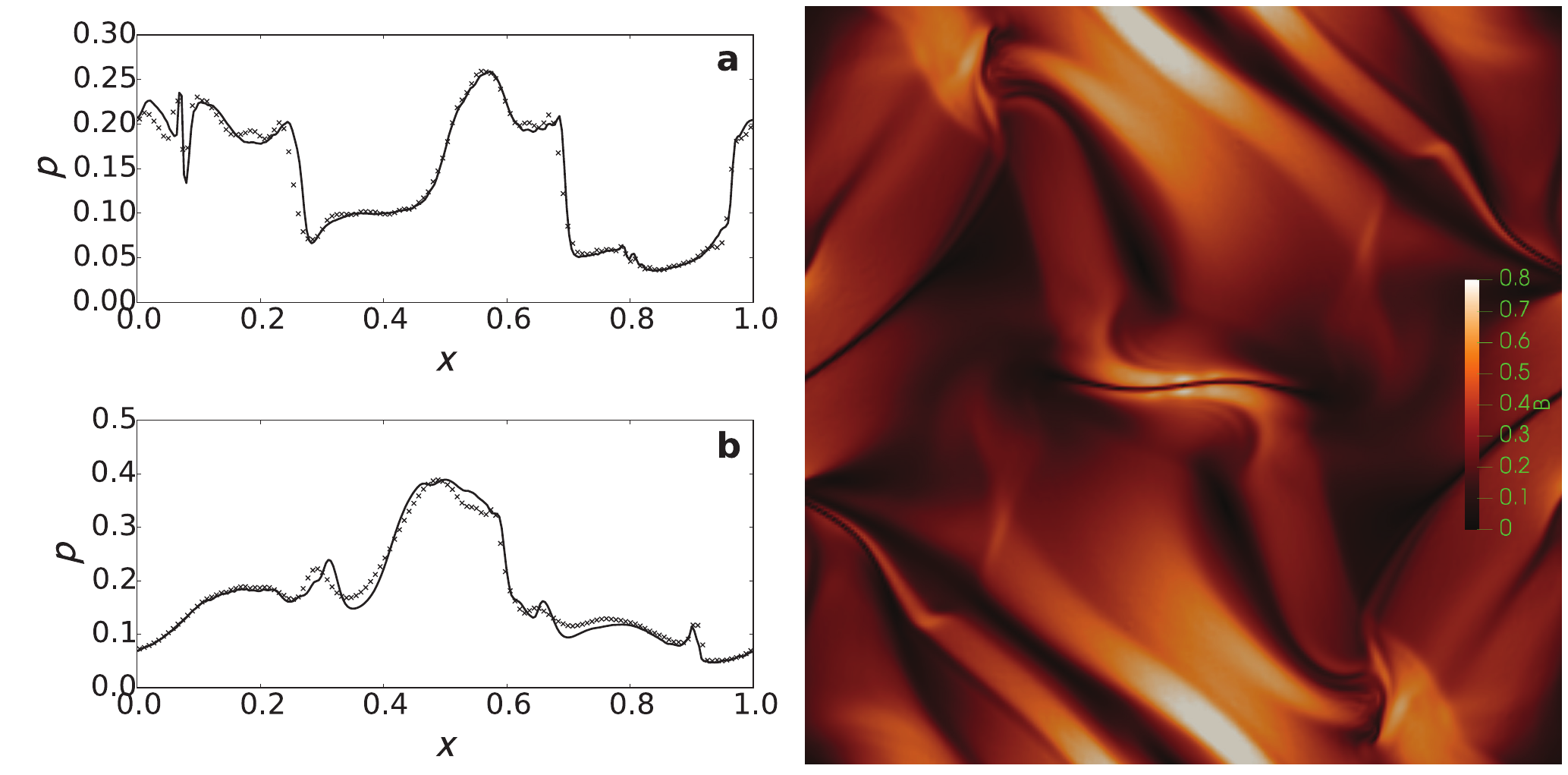}
\caption{Orszag-Tang vortex simulation at $t=0.5$. Left: pressure along two cuts, at $y=0.3125$ (a) and at $y=0.4277$ (b) in the simulations with $256\times256$ cells (solid) and $128\times128$ cells (crosses).
Right: magnetic field amplitude.
This figure may be compared with Figures~5 and 6 in \citet{Stasyszyn:etal:2013MNRAS}, and Figure~23 in \citet{Stone:2008}.
}
\label{fig:orszag-tang}
\end{centering}
\end{figure}
Orszag-Tang vortex represents a complex interaction of different MHD shocks in 2D, with a translation to MHD turbulence, and is often used as a reference test case for MHD code validation.
Initial pressure and density in a periodic domain with $L_x=L_y=1$ are uniform, $\rho=25/\left(36\pi\right)$, and $p=5/\left(12\pi\right)$.
Initial velocity components are $u_x=-\sin\left(2\pi y\right)$ and $u_y=\sin\left(2\pi x\right)$.
The magnetic field is given by $A_z=\left(B_0/4\pi\right)\cos\left(4\pi x\right) + \left(B_0/2\pi\right)\cos\left(2\pi y\right)$ with $B_0=1/\sqrt{4\pi}$.

Simulations at two grid resolutions, with $256\times256$ cells, $\Delta t=10^{-4}$, and with $128\times128$ cells, $\Delta t=5\cdot10^{-4}$, both using 25 particles/cell, are compared in Figure~\ref{fig:orszag-tang}.
Panels on the left represent gas pressure along two cuts, $y=0.3125$ (a) and at $y=0.4277$ (b), and may be used for quantitative comparison with reference results provided in Figure~23 of \citet{Stone:2008}, and in Figure~6 of \citet{Stasyszyn:etal:2013MNRAS}.
Slurm simulation with $256\times 256$ grid cells reproduces all major features of the reference solutions, except a few sharpest shock interfaces.
This is also confirmed by a snapshot of the magnetic field amplitude $B$ shown in the right panel of Fig.~\ref{fig:orszag-tang}.
This image could be compared with Fig.~5 in \citet{Stasyszyn:etal:2013MNRAS}.

\subsection{Rayleigh-Taylor}
\begin{figure}
\begin{centering}
\includegraphics[width=0.99\textwidth]{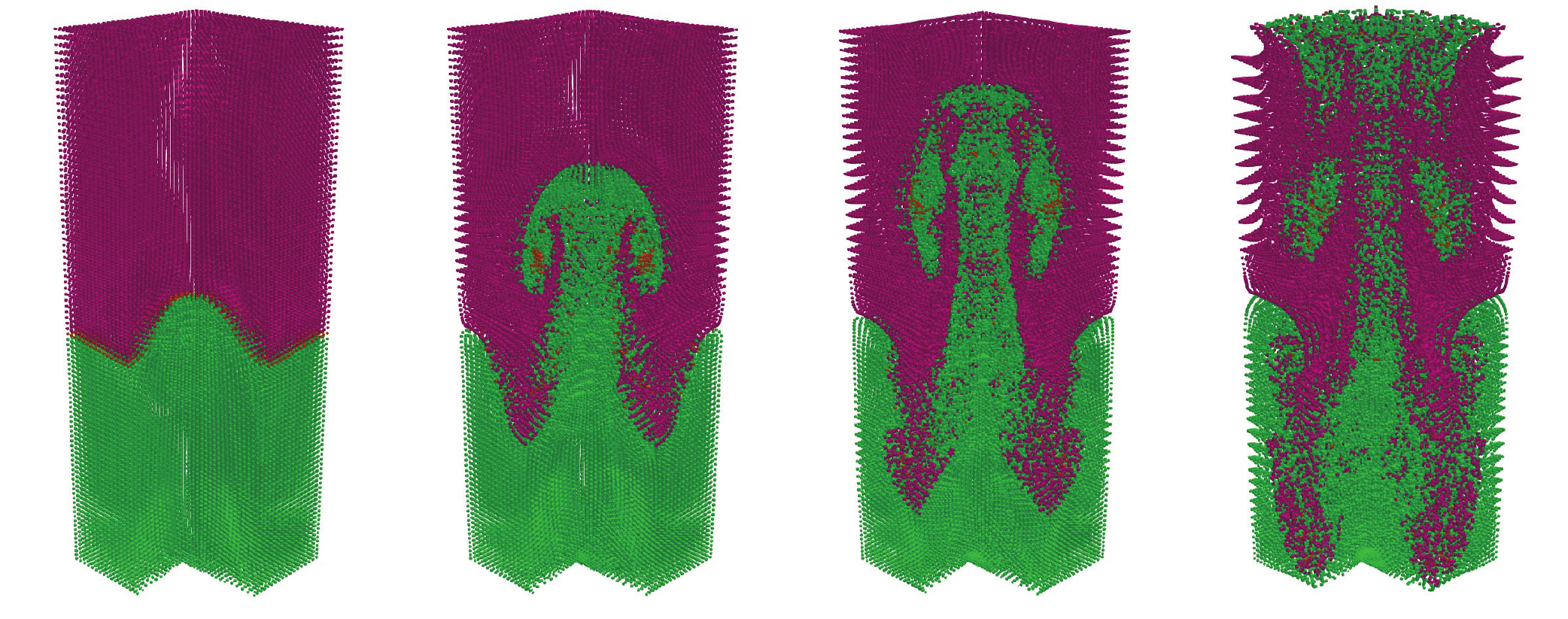}
\caption{Snapshots of Rayleigh-Taylor instability simulation at time moments $t=5$, $t=10$, $t=12.5$, and $t=15$.
Each little cube corresponds to a computational particle (only $1$ out of $27$ particles are saved).
Purple particles represent the heavy fluid, initially on top of the lighter fluid (green particles), and completely mixed at the end of the simulation.
The $3/4$ of the box in front of each snapshot have been clipped to visualize the motion inside the domain.
}
\label{fig:rayleigh-taylor}
\end{centering}
\end{figure}
A Rayleigh-Taylor instability, together with Kelvin-Helmholtz instability are the key ingredients in fluid modeling of many astrophysical phenomena, from solar convection to the solar wind-magnetosphere interaction.
As shown by \citet{Bacchini:2017}, particle volume evolution enables Slurm to successfully model the Kelvin-Helmholtz instability, even with explicit time-stepping.
Rayleigh-Taylor instability simulated by different numerical codes was compared by \cite{Liska:Wendroff:2003}.
There is no unique solution to the problem in the nonlinear regime, therefore we do not attempt to reproduce results of other codes.
Instead, we use the three-dimensional Rayleigh-Taylor problem to test the reflecting boundaries of our code, the symmetry of the solution, and demonstrate its unique capability to study fluid mixing.
No pure grid-based code is able to track individual fluid particles and thus be self-consistently used to investigate mixing of different fluids or plasmas.

The computational domain is a $0.5\times0.5\times1.5$ box with periodic X and Y boundaries, and reflective Z boundaries.
In the top half of the box the fluid is heavier with $\rho=2$, while in the bottom half $\rho=1$.
Gravity acceleration $g_z=-0.1$ and specific heats ratio $\gamma=1.4$ are constant throughout the domain.
The initial pressure is given by the hydrostatic equilibrium $p(z)=2.5+g_z\rho z$.
The instability is excited by a single-mode velocity perturbation of the form
\begin{equation}
\nonumber
u_z=u_0\frac{\left[1 + \cos\left(4\pi\left(x - 0.25\right)\right)\right] \left[1 + \cos\left(4\pi\left(y - 0.25\right)\right)\right] \left[1 + \cos\left(3\pi\left(z - 0.75\right)\right)\right]}{8},
\end{equation}
where $u_0=0.01$.

Figure~\ref{fig:rayleigh-taylor} shows the nonlinear evolution of the instability simulated in a $32\times32\times96$ grid with $27$ particles/cell and $\Delta t=10^{-3}$.
The flow exhibits classical features of the Rayleigh-Taylor instability with secondary Kelvin-Helmholtz instabilities and growing turbulence on the edges of the flow.
Thanks to the computational particles we can catch these features even at rather low grid resolution, and also track how initially different fluids mix together.

\subsection{3D magnetic loop}
\begin{figure}
\begin{centering}
\includegraphics[width=0.99\textwidth]{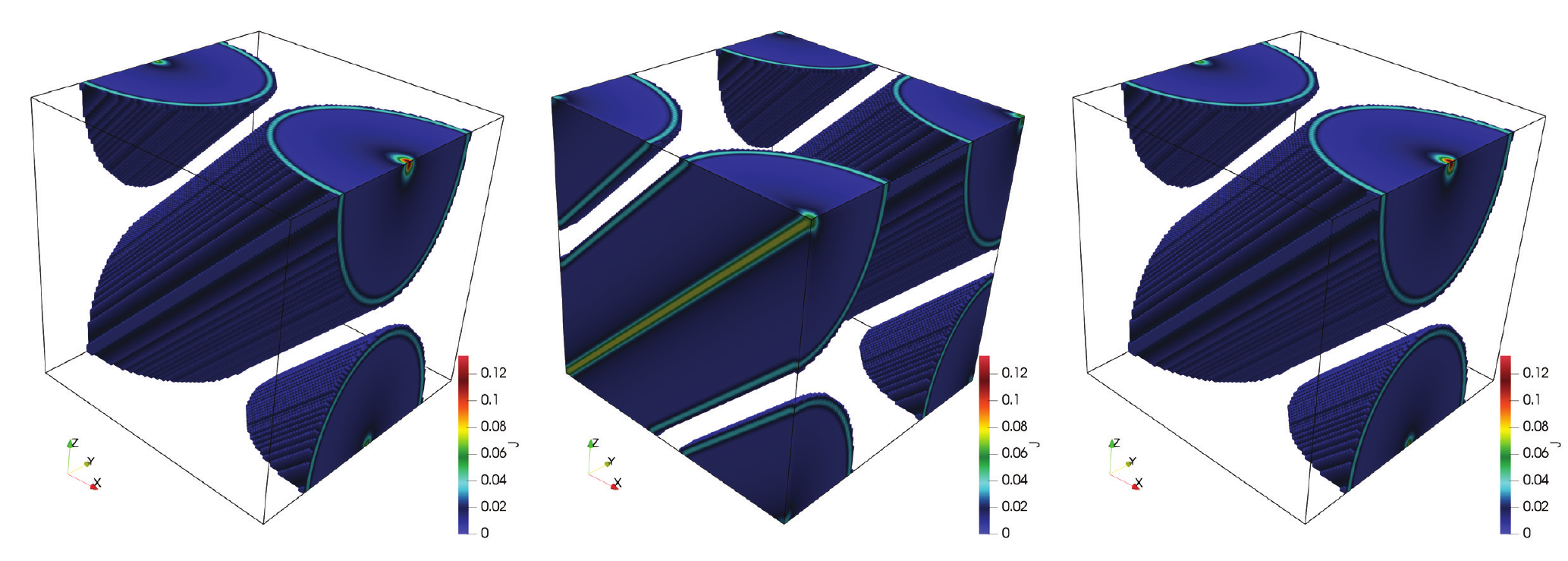}
\caption{Current density magnitude $J=\left|\nabla\times\mathbf{B}\right|$ in the beginning (left), during boundary crossing (in the middle), and after 2 crossings.
}
\label{fig:ML3D}
\end{centering}
\end{figure}
As mentioned above, in 2D simulations Slurm preserves magnetic topology exactly (Eq.~\ref{eq:Az2D}), and two-dimensional magnetic loop test is only useful to track possible bugs.
Here we show that Slurm deals with a 3D magnetic loop as easy.
In this test, the electromagnetic vector potential inside a magnetic loop, inclined by $45^{\circ}$ to the vertical axis, is given by $A_x=A_z=A_0\left( r-R \right)$, and is zero outside the loop, $r>R$, where $r$ is the distance to the loop's axis, $R=0.3$ is the loop radius, and $A_0=0.001/\sqrt{2}$.
The domain's extent is $1$ in all dimensions, and the loop is advected in all three dimensions with speed $u_x=u_y=u_z=\sqrt{3}$.
Loop's magnetic field strength is very small to ensure high plasma beta and the absence of pressure imbalance effects.

The results of the simulation in a $100^3$ cells grid with $27$ particle/cell are shown in Fig.~\ref{fig:ML3D}.
The loop is formed by two current sheets: one infinitely thin (as thin as the numerical scheme allows) on the axis, and one cylindrical on the outer boundary of the loop.
The color in the figure depicts the current density $J$ in the loop.
The right snapshot, taken after two crossings, is indistinguishable from the left snapshot which is taken in the beginning of the simulation, hence Slurm perfectly preserves magnetic topology, electric current and magnetic energy of the advected loop.
This figure could be compared with Figure~34 of \citet{Stone:2008}.

\section{Summary}
We have successfully implemented and tested a particle-in-cell MHD model, Slurm.
Several features distinguish it from all previous (known) implementations of fluid PIC:
\begin{itemize}
  \item Magnetic field evolution strategy. Fluid particles carry electromagnetic potential $\mathbf{A}$ which is projected on grid nodes from the closest particles. This way, solenoidality of magnetic field is ensured to machine precision, there is no diffusion of magnetic energy and magnetic topology is preserved accurately.
  
  \item Particle volume evolution is taken into account. Volume is used to normalize the quantites which are interpolated between the grid and the particles. Volume evolution damps numerical ringing instability, and allows us to model key hydrodynamic instabilities such as Kelvin-Helmholtz and Rayleigh-Taylor.
  
  \item Efficient bulk hyper-viscosity allows to resolve shocks and accurately model reference problems.
  
  \item Particles are carried as a double-linked list of objects. Particles can efficiently be deleted and inserted on-the-fly without reducing the code's performance. This is particularly important for space weather simulations, where open boundaries are crucial.
  
  \item The code uses task-based parallel approach in which particles are split between multiple OpenMP threads operating in shared memory.
\end{itemize}

Future developments in Slurm include open boundary conditions in complex physical models such as solar wind; optimizations of parallel performance and additional parallelization for non-shared memory systems.
Slurm is fully functional and tested, and is available online at bitbucket\footnote{\href{https://bitbucket.org/volshevsky/slurm}{https://bitbucket.org/volshevsky/slurm}. The code will be made completely open, but at the moment send us an E-mail to get access to the repository.}.
Although Slurm's primary application is envisioned in space weather modeling, the range of problems for which it could be suitable include interface flows, surface flows, fluid mixing, etc..

\vspace{5mm}
\noindent{\bf Acknowledgements}

This work is conducted under the Air Force Office of Scientific Research, Air Force Materiel Command, USAF under Award No. FA9550-14-1-0375.
V.O. and F.B. are thankful to Craig DeForest for useful discussions during the 2017 AFOSR $B_z$ meeting.

\appendix

\section{Interpolation from irregular grid in 3D}
\label{app:interpolation}
The algorithm of interpolation from 8 vertices of a hexahedron (formed by the node's closest particles) onto one enclosed point (this grid node) is adopted from NASA's \href{https://www.grc.nasa.gov/WWW/winddocs/utilities/b4wind_guide/trilinear.html}{b4wind User's Guide}.
The real coordinates of the vertices of an arbitrary hexahedron should be translated into `logical' coordinates where they form a cube.
Then interpolation weights are trivial to compute.

First, number the eight vertices of the hexahedron,
\begin{lstlisting}
  x1 = x(i  , j  , k  )
  x2 = x(i+1, j  , k  )
  x3 = x(i  , j+1, k  )
  x4 = x(i+1, j+1, k  )
  x5 = x(i  , j  , k+1)
  x6 = x(i+1, j  , k+1)
  x7 = x(i  , j+1, k+1)
  x8 = x(i+1, j+1, k+1)
\end{lstlisting}
and define the following variables\footnote{in the original document, the rightmost term (the coordinate of the point to interpolate on) in the expression for \lstinline|f0| was just ``x''. Here, we denote it as ``x0'' for clarity.}
\begin{lstlisting}
  f0 = 0.125 * (x8 + x7 + x6 + x5 + x4 + x3 + x2 + x1) - x0
  f1 = 0.125 * (x8 - x7 + x6 - x5 + x4 - x3 + x2 - x1)
  f2 = 0.125 * (x8 + x7 - x6 - x5 + x4 + x3 - x2 - x1)
  f3 = 0.125 * (x8 + x7 + x6 + x5 - x4 - x3 - x2 - x1)
  f4 = 0.125 * (x8 - x7 - x6 + x5 + x4 - x3 - x2 + x1)
  f5 = 0.125 * (x8 - x7 + x6 - x5 - x4 + x3 - x2 + x1)
  f6 = 0.125 * (x8 + x7 - x6 - x5 - x4 - x3 + x2 + x1)
  f7 = 0.125 * (x8 - x7 - x6 + x5 - x4 + x3 + x2 - x1)
\end{lstlisting}
Solve the linear system for three unknowns \lstinline|a|, \lstinline|b|, \lstinline|c|\footnote{in the original document, \lstinline|c| is sometimes confused with \lstinline|g|.} using Newton's method
\begin{lstlisting}
  f = f0 + f1*a + f2*b + f3*с + f4*a*b + f5*a*с + f6*b*с + f7*a*b*c
  g = g0 + g1*a + g2*b + g3*с + g4*a*b + g5*a*с + g6*b*с + g7*a*b*c
  h = h0 + h1*a + h2*b + h3*с + h4*a*b + h5*a*с + h6*b*с + h7*a*b*c
\end{lstlisting}
where \lstinline|g0|\ldots\lstinline|g7| and \lstinline|h0|\ldots\lstinline|h7| are defined analogous to \lstinline|f0|\ldots\lstinline|f7| in two other coordinates, y and z.
We use initial guess \lstinline|a = b = c = 0|, hence \lstinline|df == -f, dg == -g, dh == -h| and
\begin{lstlisting} 
  -f0 = f1 * da + f2 * db + f3 * db
  -g0 = g1 * da + g2 * db + g3 * db
  -h0 = h1 * da + h2 * db + h3 * db
\end{lstlisting}
When the coefficients are zero, it means the hexahedron in physical space is degenerate.
Using Kramer's method this linear system is solved for \lstinline|da|, \lstinline|db|, \lstinline|dc|.

After the new values of \lstinline|a|, \lstinline|b|, \lstinline|c| have been computed, they are used to compute the new coefficients \lstinline|f0|\ldots\lstinline|f7|, \lstinline|g0|\ldots\lstinline|g7|, and \lstinline|h0|\ldots\lstinline|h7|, and proceed with the next iteration of Newton's solver.
Iterations continue until a \lstinline|da|, \lstinline|db|, and \lstinline|dc| are small enough, or until the absolute value of any of \lstinline|a|, \lstinline|b|, or \lstinline|c| exceeds the pre-defined threshold $\gtrsim1$. 
In the latter case, the method fails, and a simple non-weighted average is used to obtain the interpolated value.

Finally, the interpolation weights for eight vertices are given by
\begin{lstlisting} 
  w1 = 0.125*(1-c)*(1-b)*(1-a)
  w2 = 0.125*(1-c)*(1-b)*(1+a)
  w3 = 0.125*(1-c)*(1+b)*(1-a)
  w4 = 0.125*(1-c)*(1+b)*(1+a)
  w5 = 0.125*(1+c)*(1-b)*(1-a)
  w6 = 0.125*(1+c)*(1-b)*(1+a)
  w7 = 0.125*(1+c)*(1+b)*(1-a)
  w8 = 0.125*(1+c)*(1+b)*(1+a)
\end{lstlisting}
i.e., \lstinline|a, b, c| are the coordinates of the interpolation point in the `logical' space where eight vertices form a cube.

It is unnecessary to provide the full proof of the method here, however the following helps to understand the basic idea.
If \lstinline|a, b, c| are the unknown `logical' coordinates, the interpolated value of any quantity \lstinline|v| given on the eight vertices of the hexahedron is given by
\begin{lstlisting}
  v = \left[(1-c)*(1-b)*(1-a)*v1 + (1-c)*(1-b)*(1+a)*v2 + 
            (1-c)*(1+b)*(1-a)*v3 + (1-c)*(1+b)*(1+a)*v4 +
            (1+c)*(1-b)*(1-a)*v5 + (1+c)*(1-b)*(1+a)*v6 +
            (1+c)*(1+b)*(1-a)*v7 + (1+c)*(1+b)*(1+a)*v8] / 8
\end{lstlisting}
Define
\begin{lstlisting}
  xa = [(1-a)*x1 + (1+a)*x2] / 2
  xb = [(1-a)*x3 + (1+a)*x4] / 2
  xc = [(1-a)*x5 + (1+a)*x6] / 2
  xd = [(1-a)*x7 + (1+a)*x8] / 2
  xe = [(1-b)*xa + (1+b)*xb] / 2
  xf = [(1-b)*xc + (1+b)*xd] / 2
\end{lstlisting}
and obtain the following identity
\begin{lstlisting}
  x == [(1-c)*xe + (1+c)*xf] / 2
\end{lstlisting}
Here, make the following substitutions
\begin{lstlisting}
  x = [(1-c) * ((1-b)*xa + (1+b)*xb) + 
       (1+c) * ((1-b)*xc + (1+b)*xd)] / 4

  x = [(1-c) * ((1-b) * ((1-a)*x1 + (1+a)*x2) + 
       (1+b) * ((1-a)*x3 + (1+a)*x4)) +
       (1+c) * ((1-b) * ((1-a)*x5 + (1+a)*x6) + 
       (1+b) * ((1-a)*x7 + (1+a)*x8))] / 8

  x = [(1-c)(1-b)(1-a)*x1 + (1-c)(1-b)(1+a)*x2 +
       (1-c)(1+b)(1-a)*x3 + (1-c)(1+b)(1+a)*x4 +
       (1+c)(1-b)(1-a)*x5 + (1+c)(1-b)(1+a)*x6 +
       (1+c)(1+b)(1-a)*x7 + (1+c)(1+b)(1+a)*x8]
\end{lstlisting}
After regrouping, we obtain a nonlinear equation for the three unknowns, $a$, $b$, and $c$, 
\begin{lstlisting}
  x =  [(x8 + x7 + x6 + x5 + x4 + x3 + x2 + x1) + // f0 
      a*(x8 - x7 + x6 - x5 + x4 - x3 + x2 - x1) + // f1
      b*(x8 + x7 - x6 - x5 + x4 + x3 - x2 - x1) + // f2
      g*(x8 + x7 + x6 + x5 - x4 - x3 - x2 - x1) + // f3
    a*b*(x8 - x7 - x6 + x5 + x4 - x3 - x2 + x1) + // f4
    a*c*(x8 - x7 + x6 - x5 - x4 + x3 - x2 + x1) + // f5
    b*c*(x8 + x7 - x6 - x5 - x4 - x3 + x2 + x1) + // f6
  a*b*c*(x8 - x7 - x6 + x5 - x4 + x3 + x2 - x1)]  // f7
\end{lstlisting}
where the coefficients represent the above defined \lstinline|f0|\ldots\lstinline|f7|.



\end{document}